\documentclass[prl, twocolumn, superscriptaddress]{revtex4-1}
\usepackage[version=3]{mhchem} 

\usepackage{xr}
\usepackage[english]{babel}
\usepackage{natbib}
\usepackage{microtype}
\usepackage{hyperref}
\usepackage{MnSymbol}%
\usepackage{wasysym}%
\usepackage{graphicx}

\hypersetup{colorlinks=true,citecolor=blue}
\graphicspath{{./images/}}

\newcommand{\unit}{\ \mathrm}

\newcommand{\GrapheneJJliterature}{Heersche2007, Du2008, Miao2009, Girit2009, OAristizabal2009, Kanda2010, Choi2010, Borzenets2011, Coskun2011, Lee2011, Rickhaus2012, Popinciuc2012,Komatsu2012, Mizuno2013, Voutilainen2011, JeongChoi2011, Borzenetz2013, Choi2013}

\newcommand{\Delft}{Kavli Institute of Nanoscience, Delft University of Technology, 2600 GA Delft, The Netherlands}
\newcommand{\Leiden}{Instituut-Lorentz, Universiteit Leiden, P.O. Box 9506, 2300 RA
Leiden, The Netherlands}
\newcommand{\NIMS}{Advanced Materials Laboratory, National Institute for Materials Science, 1-1 Namiki, Tsukuba, 305-0044, Japan}
\newcommand{\Moscow}{Laboratory for Quantum Limited Devices, Physics Department,
Moscow State Pedagogical University, 29 Malaya Pirogovskaya St., Moscow, 119992, Russia}

\newcommand{\TITLE}{Ballistic Josephson junctions in edge-contacted graphene}

\begin{document}

\title{\TITLE}

\author{V.~E.~Calado}\altaffiliation{Equal contributions}\affiliation{\Delft}
\author{S.~Goswami}\altaffiliation{Equal contributions} \affiliation{\Delft}
\author{G.~Nanda}\affiliation{\Delft}
\author{M.~Diez}\affiliation{\Leiden}
\author{A.~R.~Akhmerov}\affiliation{\Delft}
\author{K.~Watanabe}\affiliation{\NIMS}
\author{T.~Taniguchi}\affiliation{\NIMS}
\author{T.~M.~Klapwijk}\affiliation{\Delft}\affiliation{\Moscow}
\author{L.~M.~K.~Vandersypen}\email{L.M.K.Vandersypen@tudelft.nl}\affiliation{\Delft}

\date{\today}
\maketitle

\textbf{Hybrid graphene-superconductor devices have attracted much attention since the early days of graphene research\cite{\GrapheneJJliterature}. So far, these studies have been limited to the case of diffusive transport through graphene with poorly defined and modest quality graphene-superconductor interfaces, usually combined with small critical magnetic fields of the superconducting electrodes. Here we report graphene based Josephson junctions with one-dimensional edge contacts\cite{Wang2013} of Molybdenum Rhenium. The contacts exhibit a well defined, transparent interface to the graphene, have a critical magnetic field of 8 Tesla at 4 Kelvin and the graphene has a high quality due to its encapsulation in hexagonal boron nitride\cite{Dean2010,Wang2013}. This allows us to study and exploit graphene Josephson junctions in a new regime, characterized by ballistic transport. We find that the critical current oscillates with the carrier density due to phase coherent interference of the electrons and holes that carry the supercurrent caused by the formation of a Fabry-P\'{e}rot cavity. Furthermore, relatively large supercurrents are observed over unprecedented long distances of up to 1.5 $\mu$m. Finally, in the quantum Hall regime we observe broken symmetry states while the contacts remain superconducting. These achievements open up new avenues to exploit the Dirac nature of graphene in interaction with the superconducting state.}

\setcounter{table}{0}
\renewcommand{\thetable}{\arabic{table}}%
\setcounter{figure}{0}
\renewcommand{\thefigure}{\arabic{figure}}%

In particular, the chiral nature of the charge carriers in graphene is predicted to give rise to specular Andreev reflection~\cite{Beenakker2006}, and the conventional quantum Hall effect can be markedly different due to the interaction between edge states and the superconductor\cite{Hoppe2000,Chtchelkatchev2007}. Such systems also provide a unique way to probe valley-polarized edge states\cite{Akhmerov2007}, topological confinement in bilayer graphene~\cite{Martin2008a}, the interplay between superconductivity and quantum confinement or ballistic two-dimensional Josephson junctions and their response to phase coherent interference effects.

There are two important prerequisites that must be satisfied in order to observe any of these phenomena experimentally. First, the graphene-superconductor interface should be transparent and well defined. Secondly, the graphene must be of high electronic quality. In addition, for some of the above effects, a superconductor with a large upper critical field, $H_{c2}$, is required. While significant technological progress has been made in improving the quality of graphene by either suspending graphene\cite{Du2008a} or encapsulating it in hexagonal boron nitride (hBN) \cite{Dean2010,Wang2013}, the main challenge has been to combine such low-scattering graphene with a (large $H_{c2}$) superconductor. All reports on graphene-superconductor devices to date involved superconducting contacts deposited directly on the graphene surface, and diffusive transport through the device. In addition to the modest electronic quality, the use of top contacts leaves ambiguity in where exactly Andreev reflection takes place and under what spectral conditions. I.e. it is not clear how far electrons travel beneath the contact before entering the superconductor.


 \begin{figure}[!t]
\centering
\includegraphics{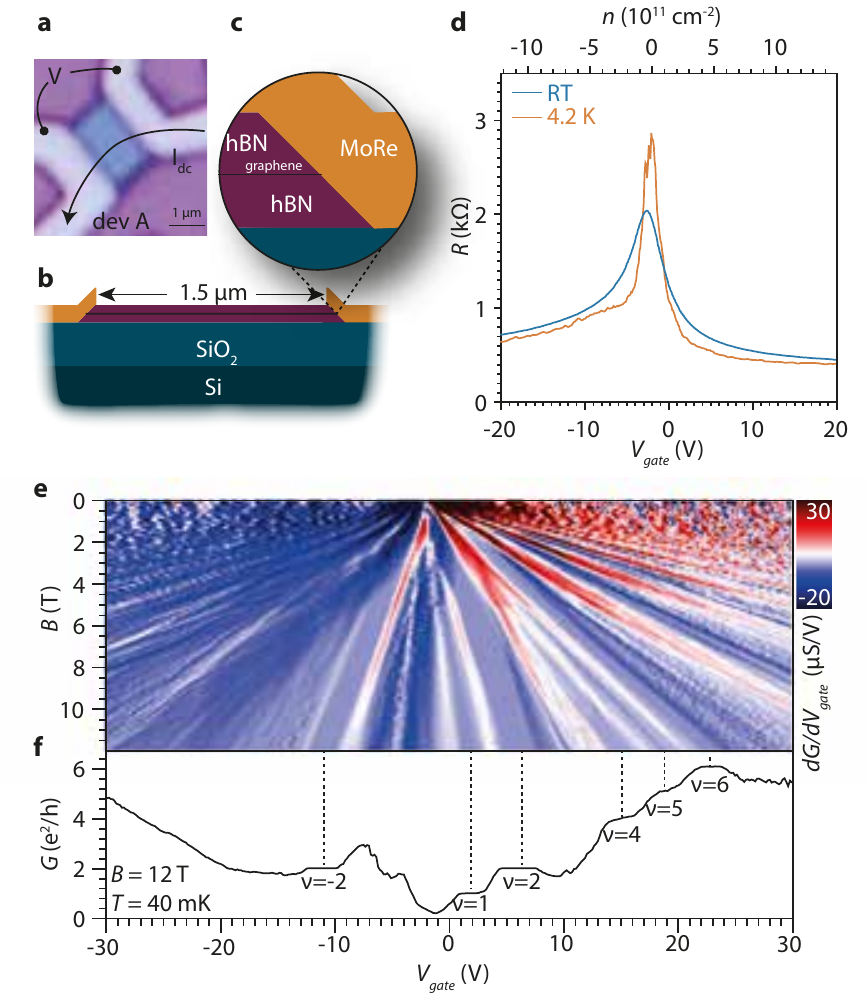}%
\caption{\textbf{High-quality hBN-Graphene-hBN devices.} \textbf{a.} An optical image of device $A$. A graphene/hBN sandwich (blue) is contacted on both sides from the edge with MoRe contacts (yellow). The contacts are split further in two, which allows a (quasi-) four probe measurement with minimal series lead resistance. \textbf{b-c.} A schematic cross-section of the device. \textbf{d.} The measured resistance, $R$, as a function of gate voltage, $V_{gate}$, at room temperature and at 4.2K. The carrier density, $n$, is extracted from Shubnikov-de-Haas oscillations. \textbf{e.} Differentiated conductance, $dG/dV_{gate}$, as a function of gate voltage and magnetic field, taken at 40 mK. \textbf{f} The conductance, $G$, as a function of gate voltage at $B = 12\unit{T}$ and $T= 40$ mK, showing the symmetry broken states.\label{fig:DC}}
\end{figure}

To realise high quality graphene-superconductor junctions we encapsulate graphene between two hBN crystals using the van der Waals pick-up method \cite{Wang2013}. This method ensures that the graphene is never in contact with any polymer during the stacking and thereafter. Electrical contact is made by metal deposition onto areas where the stack has been etched through. Unlike earlier work\cite{Wang2013}, where metal deposition is done in a separate lithography step, we start by etching only the region to be contacted, followed immediately by metal deposition. This has the following advantages: (i) our contacts are self-aligned, thereby minimizing redundant metal overlap above the graphene and reducing the screening of electric and magnetic fields and (ii) combining the etching and deposition in one step minimizes resist residues at the contact interface, which is necessary for transparent contacts. Instead of a normal metal, we sputter an alloy superconductor MoRe, which is attractive in several respects. First, MoRe is a type-II superconductor with a critical temperature $T_c\approx 8$~K and an upper critical field $H_{c2}\approx 8$~T (at 4.2~K), which should easily allow for the observation of quantum Hall states while the MoRe remains predominantly superconducting. Secondly, it has been shown that MoRe makes good electrical contact to carbon-based materials such as carbon nanotubes~\cite{Schneider2012}. Considering the fact that edge-contact resistance can vary by an order of magnitude depending on the choice of metal~\cite{Wang2013}, it is critical to select a superconductor which makes good electrical contact to graphene. This is particularly important in the context of superconductor (S) graphene (G) JJs, where the transparency of the S-G interface directly affects the Andreev reflection. Furthermore, unlike surface contacts, such one-dimensional edge contacts ensure that the Andreev reflection occurs at a well-defined location, at the edge of the graphene, where it contacts with the 3-dimensional bulk superconductor. After the deposition of the superconducting electrodes, we etch the stack into the desired geometry.

An optical image and a cross-sectional schematic of device $A$ are shown in Fig.~\ref{fig:DC}a-c. The graphene is etched to a $L=1.5\unit{\mu m}$ long and $W=2.0\unit{\mu m}$ wide rectangle, with MoRe edge contacts on either side. All measurements described here are performed in a (dc) four point geometry, as shown in Fig.~\ref{fig:DC}a. The MoRe leads are arranged such that the lead series resistance is minimized and the measured resistance is effectively the two-probe graphene resistance, irrespective of whether the MoRe is normal or superconducting. This is important since disordered superconductors such as MoRe have a large normal-state resistivity, potentially confusing the interpretation of the measurements when the electrodes turn normal (see SI).

Fig.~\ref{fig:DC}d shows the measured resistance, $R$, versus back gate voltage, $V_{gate}$, at room temperature and 4.2~K. A clear electron-hole asymmetry is visible with the resistance in the hole doped ($p$) regime being somewhat larger than that in the electron doped ($n$) regime. We attribute this to contact-induced $n$-type doping, which leads to the formation of $pn$ junctions close to the contacts when the bulk of the graphene is $p$ doped. Such $n-$type doping effects from normal edge contacts have also recently been reported \cite{Maher2014}. Fig.~\ref{fig:DC}e shows the Landau fan diagram recorded up to $B=12\unit{T}$. The high electronic quality of the graphene is evident from the emergence of broken symmetry states above $B=5\unit{T}$, which are well developed at $B=12\unit{T}$ (Fig.~\ref{fig:DC}f). To our knowledge, this is the first observation of broken symmetry states in graphene with superconducting contacts. The plateaus on the electron side are better developed than those on the hole side, presumably a consequence of doping near the contacts.
\begin{figure}[!b]
\centering
\includegraphics{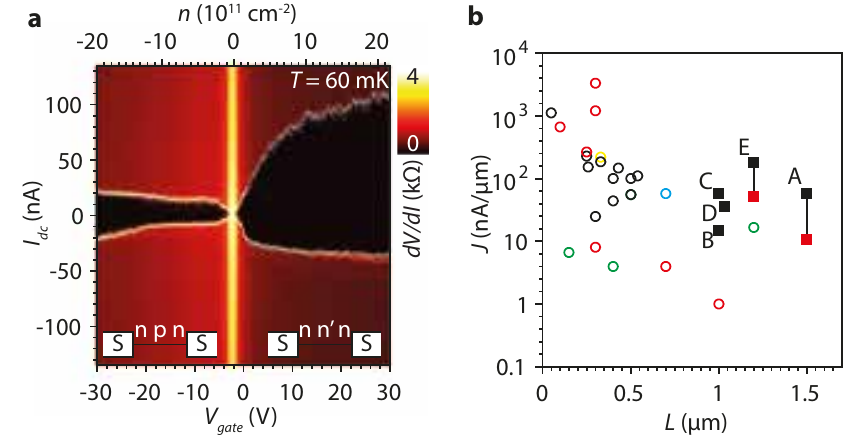}%
\caption{\textbf{Long distance Josephson current in edge-contacted graphene}. \textbf{a} The differential resistance, $dV/dI$, is plotted as a function of applied dc current bias, $I_{dc}$, and gate voltage, $V_{gate}$, at $60\unit{mK}$. \textbf{b} Critical current density, $J$, plotted as a function of device length, $L$. Squares are the side contacted MoRe graphene devices $A-E$ reported here. Black (red) squares correspond to a temperature of 50~mK (700~mK). More details about the temperature dependence can be found in the SI. Circles are data points taken from the literature \cite{\GrapheneJJliterature}. Colors indicate different superconductors used: Black circles refer to Al, green circles to Nb/NbN/NbTiN, blue ReW, red Pb/PbIn and yellow Pt/Ta.\label{fig:SC}}
\end{figure}

\begin{figure*}[!t]
\centering
\includegraphics{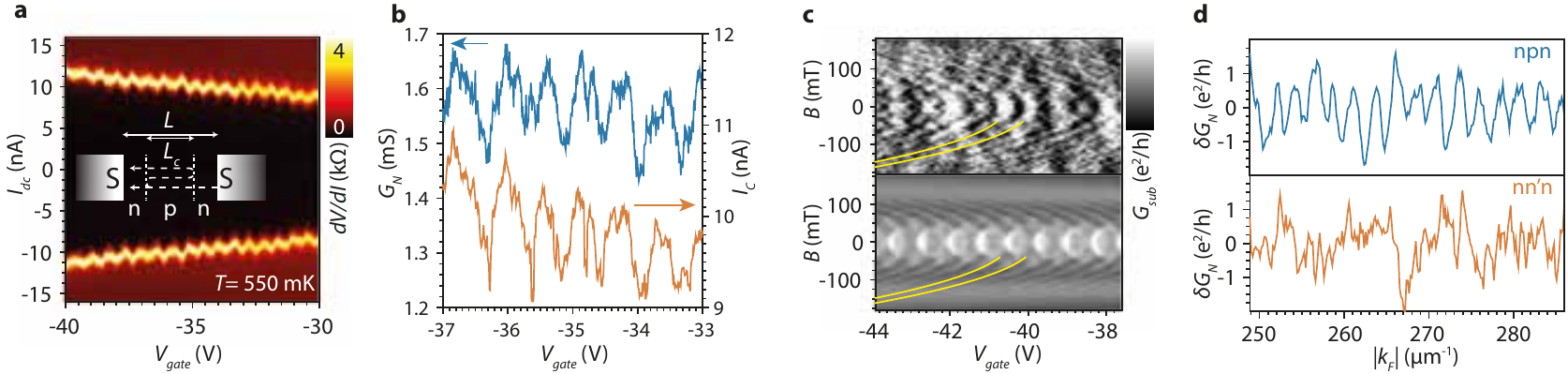}%
\caption{\textbf{Fabry P\'{e}rot resonances in a Josephson junction.} \textbf{a.} The differential resistance, $dV/dI$, is plotted as a function of the applied dc current bias, $I_{dc}$, and gate voltage, $V_{gate}$, at $T=550\unit{mK}$. At $60\unit{mK}$, statistical fluctuations in $I_{c}$ make the effect much less visible. Inset:  A schematic of a cavity formed between $pn$-junctions due to doping near the contacts. Interference occurs due to reflections at the $pn$-junctions. \textbf{b.} The normal state conductance, $G_N$, and the critical current, $I_{c}$, plotted as a function of gate voltage, $V_{gate}$. \textbf{c.} (upper panel) The conductance measured as a function of the magnetic field and gate voltage with $I_{dc}=100\unit{nA}$. The dispersion of the Fabry-P\'{e}rot interferences follows a $\mathrm{4^{th}}$-order polynomial, see equation~\ref{eq:FPdispersion}, plotted in yellow. (lower panel) The simulated conductance for a cavity size of $L=1.3\unit{\mu m}$ and $W=2\unit{\mu m}$. \textbf{d.}(upper panel) The conductance $\delta G$ in the $npn$ regime as function of absolute wavenumber $|k_F|$ in the central part of the device ($|k_F|$ is determined from the carrier density). $\delta G_N$ is obtained after subtraction of a slowly varying background (see SI for details). In the lower panel we plot $\delta G_N$ in the $nn'n$ regime for the same wavenumber range. Here we attribute the fluctuations to UCF. \label{fig:FP}}
\end{figure*}

At zero magnetic field we observe a gate-tunable supercurrent through the device. In Fig.~\ref{fig:SC}a we plot the diffential resistance, $dV/dI$, as a function of gate voltage, $V_{gate}$, and the current bias, $I_{dc}$. Evidently, the critical current, $I_{c}$, vanishes at the charge neutrality point, but reaches values in excess of 100~nA at V$_{gate}=30$~V. The individual $I_{dc}-V$ curves are hysteretic, as is evident from the asymmetry about $I_{dc}=0$ (we discuss the possible origins of this hysteresis in the SI). On the hole side $I_{c}$ is considerably smaller, consistent with the formation of the conjectured $npn$ junctions. In Fig.~\ref{fig:SC}b we plot the critical current density per unit length, $J$, versus the JJ length, $L$, with data obtained from previous reports of graphene JJs (circles) along with the present MoRe edge-contacted devices (squares). The black squares show the critical current density at $50\unit{mK}$, whereas the red squares are taken at $700\unit{mK}$. We point out that the critical current density depends on the temperature and the graphene carrier density, which vary from study to study. Despite this, it is clear that our MoRe edge-contacted devices stand out in relative magnitude compared to the previous data. We find large supercurrent densities (up to $\sim200\unit{nA/\mu m}$) over significantly longer distances ($\sim1.5\unit{\mu m}$). The observation of large supercurrents over an unprecedented long distance of 1.5~$\mu m$ indicates a high quality of both the graphene itself and of the 1D graphene-superconductor interfaces.

In addition, we find unambiguous signatures of ballistic Josephson transport in this 2D geometry. As shown in Fig.~\ref{fig:FP}a, we observe for the first time clear oscillations in the critical current and the retrapping current when we vary the gate voltage, indicative  of Fabry-P\'{e}rot (FP) interferences in the supercurrent through the junction.  The transmission probability of electrons and holes that carry the supercurrent is the result of interference of trajectories that travel ballistically from one contact to the other with multiple reflections close to or at the edges of the graphene flake. As the gate voltage is varied, the Fermi wavelength changes, constructive and destructive interference alternate, leading to modulations in the critical current. One may expect the graphene-superconductor interfaces to form the walls of the cavity. However, we observe the $I_c$ oscillations only on the hole-doped side and not on the electron-doped side (see Fig. ~\ref{fig:FP}d and the discussion below). This suggests that in the presence of $n$-doped regions near the MoRe-graphene interface the relevant cavity is instead formed by $pn$ junctions near the contacts (see inset Fig. ~\ref{fig:FP}a). This gives rise to a reduced cavity length $L_c$. This length can be directly inferred from the period of the oscillations, extracted via a Fourier analysis (see SI) of these oscillations over many periods. A cavity length of $L_c=1.3\unit{\mu m}$ is found, which is smaller than the etched device length ($L=1.5\unit{\mu m}$). A similar difference between device size and inferred cavity length was seen in device $D$ (see SI). This difference may arise from screening of the back gate near the contacts in combination with the presence of the n-doped regions at the MoRe-graphene interfaces in both devices.

The interpretation of the oscillations in $I_c$ in terms of FP interference, is further supported by comparing them with  the oscillations in the normal state conductance, $G_N$, measured at currents just above $I_c$. The oscillations of $I_c$ with gate voltage clearly match the oscillations in $G_N$, (see Fig.~\ref{fig:FP}b), as expected for Josephson junctions. In the case of normal state transport, we can apply a weak magnetic field perpendicular to the graphene, to apply a Lorentz force to the trajectories of electrons and holes. This is expected to give a characteristic shift of the FP resonances due to the accumulation of extra field-dependent phases. Indeed in the measurements shown in Fig.~\ref{fig:FP}c, we find that as B increases the main resonance features shift to higher density, following a characteristic dispersion. To enhance the visibility, we plot the quantity $G_{sub}$, which was obtained after subtracting a gate dependent (but field independent) modulation of the background conductance. We compare the data with the results of numerical simulations of the device conductance (see methods for further details) in the ballistic regime and with $npn$ junctions for the exact geometry of the measured device (Fig.~\ref{fig:FP}c lower panel). Simulation and experiment show an almost identical dispersion of the FP resonances with magnetic field. It is also possible to obtain a semiclassical expression for the resonance condition (see SI) by considering all the phases accumulated in the n-region of the $npn$ junction:
\begin{equation}
\frac{L_c}{\lambda_F(V_{gate})} = n_m +\frac{1}{2} + \frac{1}{6n_m} \left(\frac{L_c^2 e B}{h}\right)^2,
\label{eq:FPdispersion}
\end{equation}
with $n_m$ a specific integer mode, $\lambda_F(V_{gate})$ the Fermi wavelength which is tuned by the backgate ($V_{gate}\sim1/\lambda_F^2$), $L_c$ the cavity size, $e$ the electron charge and $h$ Planck's constant. The yellow curves in Fig.~\ref{fig:FP}d are calculated using Eq.~\ref{eq:FPdispersion} for modes $n_m=-121,~-120$ and show an excellent agreement with the measured and simulated results. This provides strong evidence that the observed oscillations, both in $I_c$ and $G_N$, arise from Fabry-P\'{e}rot interference, which implies phase-coherent ballistic transport. While such oscillations due to FP interference have been reported before in a variety of systems including high-quality graphene with normal contacts \cite{Young2009,Varlet2014}, here we provide evidence for phase coherent FP interference in the supercurrent, which has not been observed before in any 2D geometry.

In order to better understand the microscopic details of our device, we compare the conductance in the $npn$ regime with that in the $nn'n$ regime (Fig.~\ref{fig:FP}d). Whereas in the $npn$ regime (upper panel), we observe periodic oscillations as a function of absolute wave number, $|k_{F}|$, we observe universal conductance fluctuations (UCF) in the $nn'n$ case (lower panel). We attribute these fluctuations to  diffuse boundary scattering at or close to the graphene-MoRe interface. This diffuse scattering should also be present on the hole-side but does not dominate the transport due to the presence of the $pn$ junctions. Using the ballistic limit, $L$ much larger than the mean free path, where all resistance is from the contact interface, we can estimate a lower bound on the contact transparency, $T$ via $G = \frac{T}{2}\frac{4e^2}{\pi h}k_FW$. From the conductance in the $nn'n$ regime (see SI) we find a contact transparency of $T>0.2$. In the $npn$ case, the conductance is dominated by the $pn$ barriers. In this case, we can estimate the sharpness, $d$, of the $p$ to $n$ transition regions via $G_{npn}=\frac{e^2}{\pi h}\sqrt{\frac{k_F}{d}}W$. We find a sharpness of $d\sim70\unit{nm}$, which is a plausible value considering the device dimensions.

\begin{figure}[!t]
\centering
\includegraphics{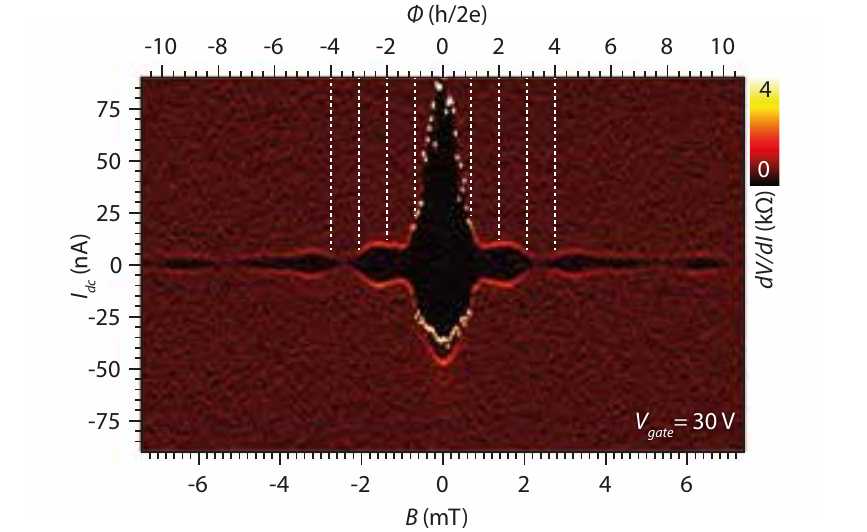}%
\caption{\textbf{Anomalous Fraunhofer diffraction pattern.} \textbf{a.} The differential resistance ($dV/dI$) is plotted as a function of applied current bias, $I_{dc}$, and magnetic field, $B$, at a gate voltage of $30\unit{V}$. We observe a a separation between minima that clearly exceeds the flux quantum, $h/2e$.
\label{fig:FH_wide}}%
\end{figure}

Since the DC Josephson effect is observed in these graphene devices over micron scale distances, we can also explore the magnetic field dependence of the critical current for unusual geometries. Earlier reports concerned graphene Josephson junctions with lengths much shorter than their width. In this case, the magnetic field dependence of $I_c$ is expected to follow the standard Fraunhofer diffraction pattern observed in tunnel junctions\cite{Tinkham}. In the present devices, in contrast, the aspect ratio is close to 1, which has two consequences. First, unlike in tunnel junctions, the phase difference across the junction must be integrated along both interfaces. Furthermore, contributions involving reflections off the side of the junction must be included, especially when transport is ballistic\cite{Heida1998, Ledermann1999, Sheehy2003}. The main prediction in this case is that the periodicity of $I_c$ with magnetic flux becomes larger than a single flux quantum, $\Phi_0 = h/2e$. Despite significant differences across the patterns measured on the various devices, we consistently find a period larger than $\Phi_0$, as seen in Fig.~\ref{fig:FH_wide} for device $A$ (and in the SI for devices $B$ and $C$). In contrast, earlier reports on graphene Josephson junctions all show flux periods smaller than $\Phi_0$ \cite{Heersche2007, Du2008, OAristizabal2009, Choi2010, JeongChoi2011, Coskun2011, Popinciuc2012, Komatsu2012} before corrections to account for the London penetration depth.

The Fabry P\'{e}rot oscillations in the critical current and the anomalous Fraunhofer diffraction patterns are mutually consistent and provide strong evidence of ballistic effects in superconducting transport through graphene. We believe that this is the first unambiguous demonstration of a ballistic JJ in graphene.

\subsection{Methods}
\subsubsection{DC transport measurements}
Low temperature dc measurements are done in a Leiden Cryogenics MCK-50 $\mathrm{^3He/^4He}$ dilution fridge. The setup can reach a base temperature of 40 mK with an electron temperature of about 70 mK. DC Currents and voltages are applied and probed with a home-built measurement setup. Furthermore the setup is equipped with a superconducting magnet coil that can sustain fields of up to 12 T.
\subsubsection{Tight-binding simulation}
The FP oscillations in the $npn$ junction are simulated by a tight-binding calculation using the Kwant software package\cite{Groth2014}. The source code has been provided along with this submission as an ancillary file. A $1.5\unit{\mu m}\times 2.0\unit{\mu m}$ hexagonal lattice is discretized with a lattice constant of $a = 2\unit{nm}$, with metallic leads on the $2.0\unit{\mu m}$ wide sides. The contact induced doping near both leads is modeled by a $100\unit{nm}$ region with a fixed chemical potential. The width of the transition region from the $n$ to the central $p$ region is set to $50\unit{nm}$ and modelled by $\tanh{\left[(x-x_0)/25\unit{nm}\right]}$. A finite contact resistance is imposed by reducing the transparency between the central strip and the leads to $60\%$. Finally we calculate the transmission as a function of the Fermi wavenumber $k_F(\mu_p)$ and magnetic field $B$, resulting in the dispersion given in Fig.~\ref{fig:FP}d.

\subsection{Acknowledgements}
We thank Vibhor Singh for sharing the MoRe sputtering recipe and Carlo Beenakker for fruitful discussions. We acknowledge support from the EC-FET Graphene Flagship, from the European Research Council Advanced grant No.~339306 (METIQUM),  from a European Research Council Synergy grant (QC-LAB), and from the Ministry of Education and Science of the Russian Federation under Contract No.~14.B25.31.0007. This work is part of the Nanofront consortium, funded by the Dutch Science Foundation OCW/NWO/FOM.

\subsection{Author contributions}
K.W. and T.T. grew the hBN crystals, G.N. fabricated the devices, V.E.C. and S.G. performed the measurements, and M.D. and A.R.A. did the numerical simulations and theory. The measurements were analyzed and interpreted by V.E.C, S.G., M.D, A.R.A, T.M.K., and L.M.K.V. The manuscript was written by V.E.C, S.G. and L.M.K.V. with input from A.R.A., M.D. and T.M.K.

The authors declare no competing financial interests.

\clearpage
\setcounter{figure}{0}
\renewcommand{\thefigure}{S\arabic{figure}}

\section{\large{Supplementary Information}}

\author{V.~E.~Calado}
\author{S.~Goswami}
\author{G.~Nanda}
\author{M.~Diez}
\author{A.~R.~Akhmerov}
\author{K.~Watanabe}
\author{T.~Taniguchi}
\author{T.~M.~Klapwijk}
\author{L.~M.~K.~Vandersypen}

\maketitle

\begin{figure*}[!h]
\centering
\includegraphics{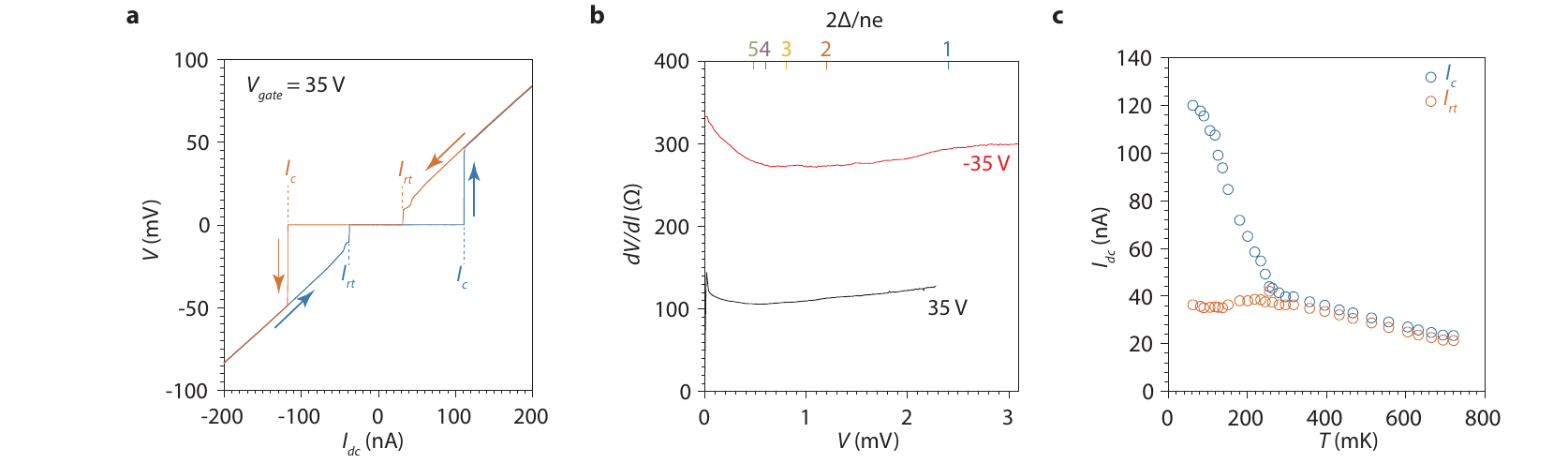}
\caption{ \textbf{Current-voltage characteristics and temperature dependence}
\textbf{a.} A current-voltage curve (in device A) at 50~mK and large n-doping showing a characteristic hysteresis. Such hysteresis is often attributed to the existence of an underdamped Josephson junction. An alternative explanation for such hysteresis lies in non-equilibrium or heating effects in the JJ. In this scenario increasing the current results in an increased power fed through the JJ, Joule heating, and an increased electron temperature compared to the temperature of the reservoirs. Both these processes could be at play in our graphene JJs, and one would require further detailed studies to determine the origin of the hysteresis. \textbf{b.} Bias spectroscopy (device E) shows no discernable features associated with multiple Andreev reflections (MAR). Top axis shows the expected positions of these features. There can in principle be several reasons for the absence of discernable MAR features in the differential resistance, such as extremely low (or high) transparency or the presence of residual inelastic scattering. To resolve this question we need more data from a range of devices, a more accurate determination of the contact transparency, and a more mature theoretical model for ballistic graphene-superconductor devices.  \textbf{c.} Temperature dependence of $I_c$ and $I_{rt}$ for device A. We can compare this data with the behavior predicted by either the Eilenberger equations (for clean SNS junctions) or the Usadel equations (for diffusive junctions). Both models predict a smooth variation of $I_c$ with temperature, so the pronounced kink around 300~mK requires a separate explanation. \label{fig:S1}}
\end{figure*}

\section{Device fabrication}
Graphene flakes are mechanically exfoliated from natural graphite (NGS naturgraphit GmbH) with blue Nitto tape (SWT 20+ Nitto Dekko Corp.) and put on Si substrates with a 90 nm thick thermally grown SiO$_2$ top layer (IDB technologies Ltd.). Single layer graphene is located under an optical microscope (Olympus BX51).  The bottom hexagonal Boron Nitride (hBN) flakes are made in the same way as the graphene flakes, using substrates with 285 nm thermally SiO$_2$ (NOVA Electronic Materials, LLC). The top hBN flakes are prepared on polymer coated transparent substrates. The transparent substrates consist of a microscope glass slide with a $\sim 1\times 1\unit{cm^2}$ PDMS (Polydimethylsiloxane) film(GelPak, WF-40-X4-A). On top of that we spin (1500 RPM for 55 sec) 9 wt\% MMA (Methylmethacrylate) in Ethyl lactate and bake it for 20 min at 120 $\mathrm{^\circ}$C. The MMA film acts as a transfer agent and the PDMS film promotes a cushioning effect during the pickup and transfer. The glass slide with hBN flakes is mounted in a micromanipulator. The setup is similar to that of ref.~\cite{Castellanos-Gomez2014} but with an additional heating stage implemented. The graphene on Si/SiO$_2$ substrate is heated to about $80\unit{^\circ C}$ and brought in contact with the top hBN. The MMA becomes soft and the graphene is picked up through adhesion to the hBN. Subsequent transfer is done at a higher temperature of $100-150\unit{^\circ C}$. The stack is patterned using standard E-beam lithography with a thick PMMA mask and is shaped by reactive ion etching in a $20 \unit{ml~min^{-1}}$ O$\mathrm{_2}$ plasma during 1 min (Leybold-Heraeus, 60 W, $50\unit{\mu bar}$) and subsequently in a $40/4 \unit{ml~min^{-1}}$ CHF$\mathrm{_3}$ /O$\mathrm{_2}$  plasma during 1 min (60 W, $50\unit{\mu bar}$). Next 80 nm MoRe is deposited by a sputtering process using a DC plasma with a power of 100 W in Ar. Lift-off is completed in hot acetone ($T=54\unit{^\circ C}$) for about 3-4 hours. A second lithography step defines the intended graphene geometry in a PMMA / hydrogen silsesquioxane (HSQ) mask, followed by an etch step using the same plasma conditions as in the first lithography step. The fabricated devices are not exposed to any (Ar/H2) annealing processes that are common for graphene devices.

\section{Details on the $npn$-junction semiclassics}
In Fig.~\ref{fig:FP}c of the main text we have highlighted the dispersion of dominant Fabry P\'{e}rot resonances with the external magnetic field. Here we derive these resonances from the semiclassical phase accumulated in the $p$-region of the $npn$-junction.
If the junction is along the $x$-direction the total Landauer conductance of the junction can be expressed as a sum of transmission amplitudes over transverse momenta $k_y$.
The leading contribution of the oscillating part of the conductance is given by \cite{Young2009}
\begin{align}
  G_{osc} &= \frac{8e^2}{h}\sum_{k_y}|T_+|^2|T_-|^2|R_+||R_-|\cos\theta e^{-2L_{\rm c}/\ell}.
  \label{eq:Gosc}
\end{align}
Here $L_{\rm c}$ is the length of the $p$-doped region (the cavity), $\ell$ the mean-free path in the $p$-region.
Choosing the center of the junction at $x=0$ and using the Landau gauge where $\bm A=Bx\hat y$, the transmission and reflection amplitudes at the junctions located at $x=\pm L/2$ are
\begin{align}
  T_\pm &= \exp\left[-\frac{\pi\hbar v_{\rm F}}{2eE}\left( k_y\pm \frac{eBL}{2\hbar} \right)^2\right],
  \label{eq:Tpm}
\end{align}
and $|R_\pm| = \sqrt{1-|T_\pm|^2}$.
Finally
\begin{align}
  \theta_{\rm WKB} &= \int_{-L/2}^{L/2} \sqrt{k_{\rm F}^2-\left( k_{y}-\frac{e}{\hbar}Bx \right)^2}
  \label{eq:thetaWKB}
\end{align}
is the semiclassical phase difference across the cavity for neighboring trajectories.
For simplicity we have assumed that the junctions at $x \pm L/2$ are sharp.
We are interested in resonances at finite fields where the transmission is dominated by small transverse momenta (see below) which acquire an additional nontrivial $\pi$ phase, up on reflection from the $pn$-junction. The total accumulated phase across the cavity and back is $\theta=2\theta_{\rm WKB}+\pi$.
At zero magnetic field $k_y=0$ is equivalent to normal incidence yielding a perfect transmission (Eq.~\eqref{eq:Tpm}) and a vanishing contribution to the interference. In this regime the dominant contribution to the Fabry-P\'{e}rot resonances stems from finite momenta. The bending of the semiclassical trajectories due to the cyclotron motion at finite field causes a switch in the dominant contribution of the conductance to small momenta. The transition is characterized by a phase shift in the conductance oscillations at \cite{Young2009}
\begin{align}
  B^\star = \sqrt{\frac{\hbar e E}{e^2 v_F L_{c}^2}} \,.
  \label{eq:Bstar}
\end{align}

For fields $B$ larger than the characteristic field $B^\star$ the largest contribution to the total conductance Eq.~\eqref{eq:Gosc} is from $k_y=0$ ($|T_+|^2|T_-|^2|R_+||R_-|$ is maximized by $k_y=0$). The switching field dependents on the degree of collimation at the $np/pn$-junctions which is determined by the electrical field $E=V/d$, where $V=(\mu_{\rm n}-\mu_{\rm p})/e$ across the junction.
An estimate for the junction thickness $d=70\,{\rm nm}$ is given in the main text. Unfortunately it is not possible for us to get a reliable estimate of the doping level in the n-doped region near the contacts. We observe $pnp$-junction behavior up to very large negative back gate voltages. This suggest that screening from the contacts plays an important role in the local density level and profile rendering even a crude estimate on simple grounds impossible. From our experimental measurements we estimate the switching field to be below $20\,{\rm mT}$. This implies that the local electric field should be on the order of $1\,{\rm V}/{\rm \mu m}$ or smaller. The field range studied in Fig.~\ref{fig:FP}c is well above this switching field allowing us to focus on the regime where the $k_y=0$ contribution to the conductance oscillations dominates.
This contribution is resonant when the total acquired phase for $k_y=0$ is an integer multiple of $2\pi$
\begin{align}
  2\theta_{\rm WKB}|_{k_y=0} + \pi = 2\pi n.
  \label{eq:resonancecond}
\end{align}
For small fields where the cyclotron radius is larger than the junction length $L$ we can expand $\theta_{\rm WKB}$ to lowest order in $B$
\begin{align}
  \theta_{\rm WKB} \approx k_{\rm L}\left( 1-\frac{e^2L^2B^2}{24\hbar k_{\rm F}^2} \right).
  \label{eq:thetaWKBapprox}
\end{align}
Using $k_{\rm F}\sim 270\,{\rm \mu m}^{-1}$ and $B=100\,{\rm mT}$ we obtain a cyclotron radius $r_{\rm c}=\hbar k_{\rm F} /eB=1.8\,{\rm \mu m}$. This confirms that our experiments are still in this regime. Finally from the resonance condition equation~\eqref{eq:resonancecond} we find
\begin{align}
  k_{\rm F}L &= n\pi + \frac{\pi}{2} + \frac{\pi}{6n}\left(L^2\frac{e}{h}B\right)^2.
  \label{eq:resonance}
\end{align}

\begin{figure*}[t]
\centering
\includegraphics{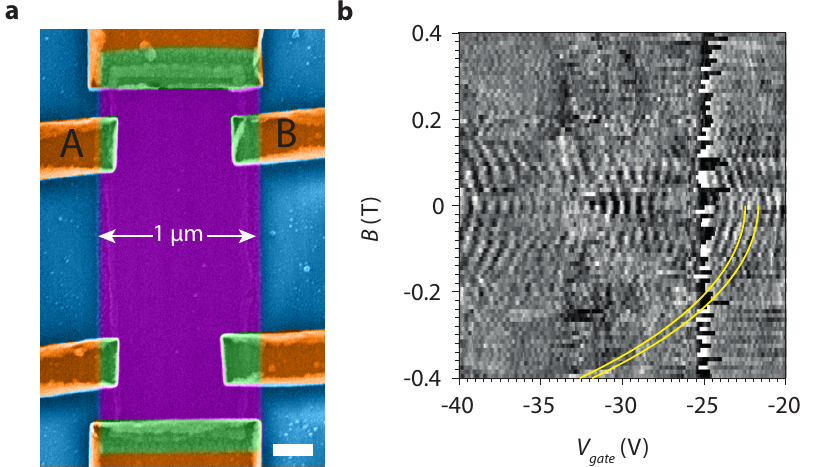}
\caption{ \textbf{Fabry-P\'{e}rot interference in device $D$.}
\textbf{a.} False colored scanning electron microscope image of device $D$ with MoRe contacts. The measurements in panel {\bf b} are recorded between the contacts indicated in the figure. The distance between the contacts is $1\unit{\mu m}$ and the contact width is $360\unit{nm}$. \textbf{b.} The conductance $G$ between the contacts indicated
in panel \textbf{a} plotted as function of magnetic field $B$ and gate voltage. Here we observe a very similar parabolic
dispersion as in Fig.~\ref{fig:FP}c in the main text. The yellow lines are calculated via
equation~\ref{eq:FPdispersion} in the main text for $L=0.8\unit{\mu m}$ and $n_m=-56,~-55$. These results imply that the transport across
the Hall bar device is ballistic. \label{fig:FP_BN18}}
\end{figure*}

\begin{figure*}[t]
\centering
\includegraphics{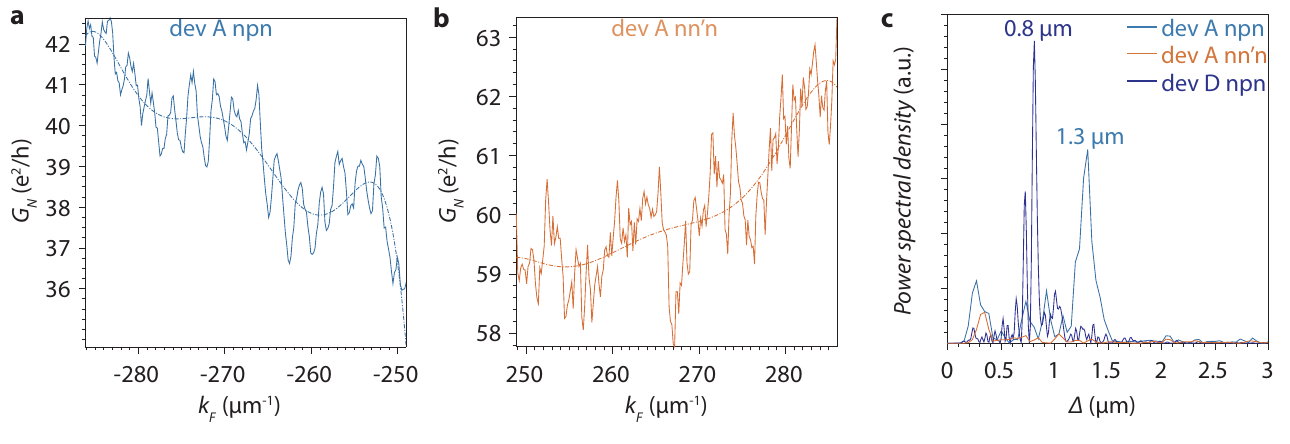}
\caption{ \textbf{Power spectral density.}
\textbf{a.} The normal state conductance $G_N$ measured in the $npn$ regime as a function of number $k_F$. The dashed line is the slowly varying background, defined here by a 6$^\mathrm{th}$-order polynomial fit.
\textbf{b.} The normal state conductance $G_N$ in the $nn'n$ regime as a function of wavenumber $k_F$, with the dashed line the slowly varying background. By taking the discrete Fourier transform $\mathcal{F}\{\delta G_N(k_F)\}=Y(\Delta)$ of the conductance fluctuations in {\bf a} and {\bf b} (after subtracting the steep background, see also Fig. \ref{fig:FP}d, we can obtain the power spectral density
$PSD(\Delta) = Y(\Delta)Y^\ast(\Delta)$. \textbf{c.} The power spectral density $PSD$ for device $A$ in the $npn$ and $nn'n$ regime and for device $D$ in the $npn$ regime as a function of length $\omega/2$.
Device $A$ and $D$ have a designed length of $L_{A}=1.5\unit{\mu m}$ and $L_{D} = 1.0\unit{\mu m}$, respectively. From the peak positions in the
$PSD$ we extract a cavity size of $L^{FP}_{A} = 1.3\unit{\mu m}$ and $L^{FP}_{D} = 0.8\unit{\mu m}$. We attribute
the difference of $\sim200\unit{nm}$  to two effects. Firstly, the size of the graphene/hBN is most likely slightly
smaller than the designed size due to the etching process. Secondly, the cavity is formed between the $pn$
junctions, where the $n$-region is caused by doping from the contact and has a finite length. This would mean the
$n$-doped region is about $\sim100\unit{nm}$ on each side. The fact that the $pn$ junctions act as the cavity walls leading to FP interference, is further supported by the fact that PSD in the $nn'n$ regime shows no predominant period. \label{fig:PowerSpecDens_NS6}}
\end{figure*}

\begin{figure*}[!h]
\centering
\includegraphics[width=.87\linewidth]{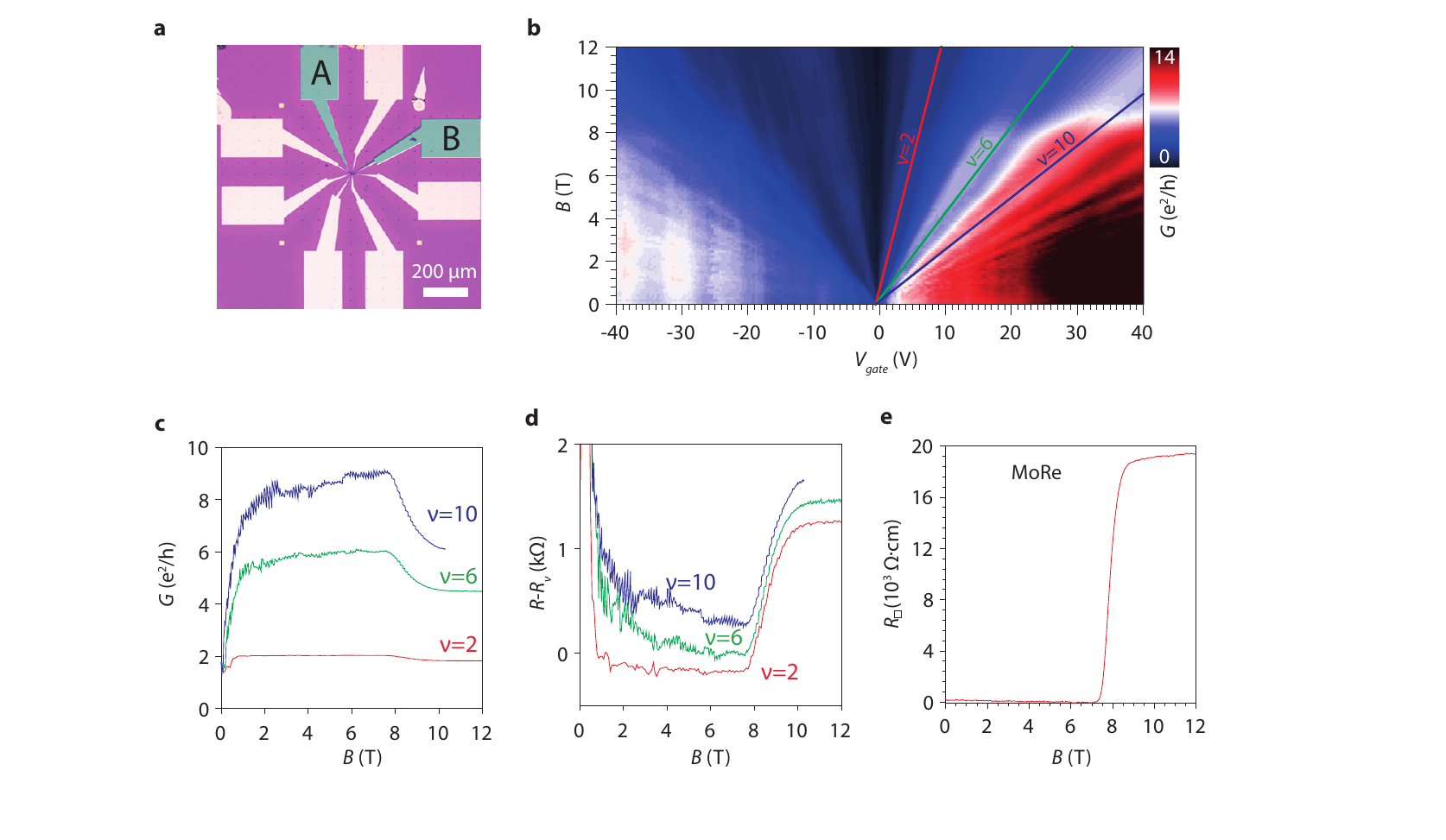}
\caption{\textbf{Lead resistance switching at 4.2 K.} \textbf{a.} An optical image of the die of device $D$. In this device, transport is measured between contacts $A$ and $B$ in a two-terminal configuration. In the normal state, the MoRe leads have a non-negligible resistance. This causes artifacts in the measurement when switching between the superconducting and normal state of the MoRe. Such artifacts from the lead resistance are minimized in the quasi four-terminal configuration used in device $A$, see Fig.~\ref{fig:DC}a in the main text. \textbf{b.} Landau fan diagram of Device $D$ (shown also in Fig~\ref{fig:FP_BN18}a), showing the conductance as a function of magnetic field $B$ and gate voltage $V_{gate}$. Above the critical field of MoRe $H_{c2}\sim 8\unit{T}$, the MoRe leads switch to the normal state, causing a jump in the two-terminal conductance. \textbf{c.} The conductance $G$ in units of $e^2/h$ for filling factors $\nu=2,~6~\mathrm{and}~10$ as a function of magnetic field $B$. \textbf{d.} When plotting resistance instead of conductance, it becomes clear that there is a fixed resistance increase when the MoRe leads turn normal. To facilitate comparison, we plot $R-R_{\nu}$, where $R_{\nu}= \frac{h}{\nu e^2}$ is the ideally expected resistance for filling factor $\nu$. \textbf{e.} The square resistance $R_\Box$ of the MoRe sheet measured using a Hall bar as a function of the magnetic field $B$. We observe again the upper critical field of MoRe of about $H_{c2}\sim 8\unit{T}$  \label{fig:QHE_switching}}
\end{figure*}

\begin{figure*}[t]
   \centering
\includegraphics{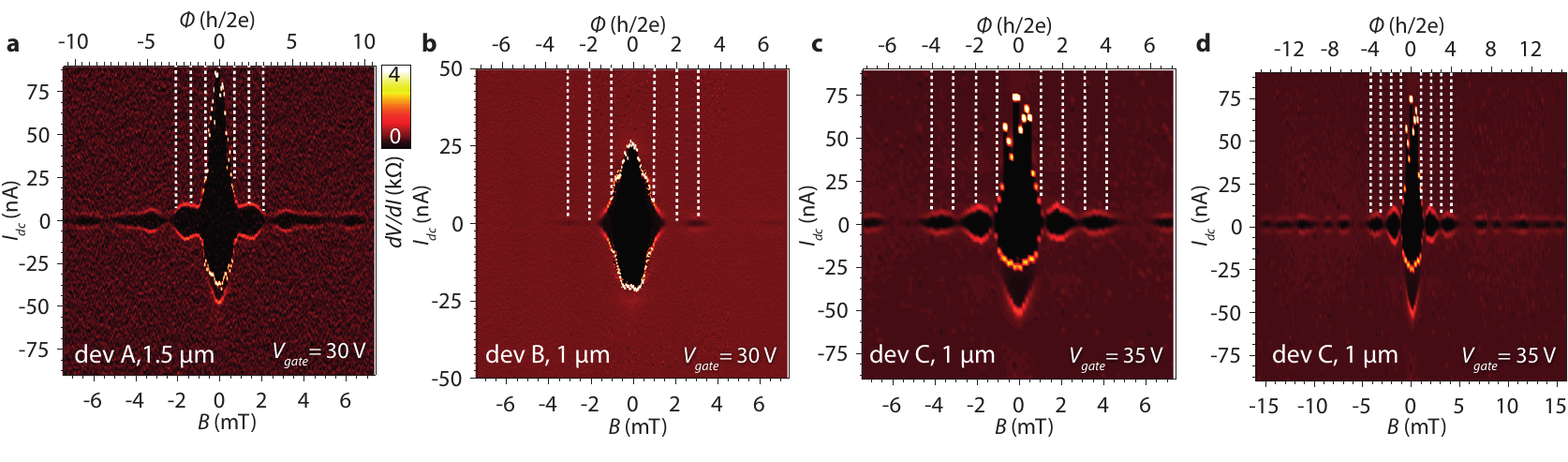}%
\caption{ \textbf{Anomalous Fraunhofer diffraction patterns.}
 \textbf{a.} The differential resistance ($dV/dI$) as a function of applied current bias, $I_{dc}$, and magnetic field, $B$ at a gate voltage of $30\unit{V}$, for device $A$. This is the same plot as in the main text, Fig.~\ref{fig:FH_wide}. \textbf{b-c.} Similar plots for devices $B$ and $C$. \textbf{d.} The same measurement as in \textbf{c}, but shown over a larger magnetic field range.
The lobes persist up to at least 20 mT which corresponds to about 9-10 periods.
In general, the dependence of $I_c$ on $B$ shows a wide range of behaviors, seen through the various Fraunhofer diffraction patterns in the four panels. An important question is what kind of variations in our samples contribute to the variation in the Fraunhofer response. For instance, we do not know a priori whether the edge contacts have uniform transmission over the full width of the contact. Despite the variations, we consistently observe that (1) when deducing the flux periodicity from the first lobe, the flux periodicity is larger than one flux quantum $\Phi_0$ and (2) the minima in $I_c$ do not reach zero between the lobes.\label{fig:FH}}%
\end{figure*}

\end{document}